\documentstyle{article}
\def\G{\Gamma}
\def\g{\gamma}

\def\p{\phi}

\def\da{\dag}

\def\a{\alpha}
\def\b{\beta}
\def\pa{\partial}

\def\l{\lambda}
\def\ba{\begin{array}}
\def\ea{\end{array}}

\def\w{\wedge}

\def\O{\Omega}

\def\ti{\times}
\def\be{\begin{equation}}
\def\ee{\end{equation}}
\def\bea{\begin{eqnarray}}
\def\eea{\end{eqnarray}}

\def\f{\frac}

\begin{document}
\thispagestyle{empty}
\large
\vskip 4mm
\rightline{  }
\vskip 1cm
\centerline{\Large \bf Two Particle Physics Models With Spontaneous $CP$ }
\vskip 0.3cm
\centerline{\Large \bf Violation From Gauge Theory On Discrete Group}

\vskip 0.9cm
\centerline{\bf Han-Ying Guo, Ke Wu, Chi Xiong}
\vskip0.5cm
\centerline{\bf 
Institute of Theoretical Physics}
\centerline{\bf 
Academia Sinica}
\centerline{\bf 
Beijing 100080,China}          
\centerline{\bf }                                                       
\vskip0.5cm
\begin{abstract}
 Based on  the differential calculus and the gauge theory on discrete groups,
  we reconstruct two physics models with spontaneous $CP$ violation: 
  (1) The Georgi-Glashow-Lee model with two Higgs triplets; 
  (2) The Weinberg-Salam-Branco model with three Higgs doublets 
  and the natural flavor conservation (NFC).
 We focus on the Lagrangian terms containing the Higgs particles and show that 
 with an appropriate choice of the discrete groups, we can obtain 
 the physically meaningful Yukawa couplings and the Higgs potentials 
 which lead to the spontaneous $CP$ violation consequently.  
\end{abstract}
\vspace{4ex}

\section{Introduction}
There were many mechanisms to explain the $CP$ nonconservation after the  
theoretical prediction raised by Lee, Oheme and Yang as early as 1956 and
the experimental support, ${K_L}^0 \rightarrow \pi^{+} \pi^{-}$ observed by
Chritenson, Cronin, Fitch and Turlay in 1964 [1]. But it  still remains an open
question for none of the explanation could match all the experimental data   
fairly well. Among these mechanisms, the spontaneous $CP$ violation (SCPV)
first suggested by T.D. Lee [1] has drawn much attention because it seems 
natural
and elegant, the total Lagrangian is assumed to satisfy invariance under $CP$
and $T$ transformation, renormalizability and invariance under certain weak-
electromagnetic gauge transformations. Lee also generalized a more realistic 
physical model at that time, Georgi-Glashow Model to a version with SCPV. In
this model, two Higgs triplets were introduced to provide SCPV and satisfy other 
physical condition. 
Having considered the natural flavor conservation
(NFC), G.Branco gave a "minimal" model based on the standard $ SU(2) \ti U(1)$ 
gauge theory with three Higgs doublets [2], the $CP$ nonconservation is entirely 
due to Higgs-boson exchange
, not in the manner of Kobayashi-Maskawa [3]. Recently, Y-L Wu proposed a 
$SU(2)_L \ti U(1)_Y $ gauge theory with two Higgs doublets which has both the
SCPV and NFC, the Glashow- Weinberg criterion for NFC is sufficient but not
necessary [4].

In these mechanisms, the Higgs particles play a key role to violate $CP$. 
But we did not understand the origin of Higgs particles very well. The 
development of non-commutative geometry (NCG) has made a great progress toward
the understanding for the origin of Higgs particles [5-9]. According to NCG, 
the Higgs fields are gauge fields with respect to discrete symmetry,  the 
Higgs potential can be viewed as Yang-Mills term in general gauge field theory
and can be determined with less ambiguity.  Moreover, the non-commutative
Yang-Mills term together with fermions can reproduce many particle physics 
models
with some constraints in free parameters. Lots of 
efforts have been made along this way.[10-18]  

 This paper is one of our series of work
on developing the gauge theory on discrete group. We start with the NCG and 
take the discrete group to be $
Z_2 \ti Z_2 $ and $ Z_2 \ti Z_2 \ti Z_2$ respectively, then construct physics 
models with SCPV.  Following the same way in  \cite{Li} \cite {Chen},
 we get the Yukawa coupling terms
, Higgs kinetic terms as well as Higgs potential which is important in SCPV.
The paper is arranged as follows: In section 2 we give a brief introduction
of gauge theory on discrete groups. In section 3 we reconstruct the  
Georgi-Glashow-Lee model in the spirit of NCG. In section 4 we take the discrete 
group to be $ Z_2 \ti Z_2 \ti Z_2$ and obtain Weinberg-Salam-Branco 
three Higgs model with SCPV, while the NFC are ensured naturally. 
 At section 5 we end with some discussions and remarks.

\section{The Gauge theory on the Discrete Group $Z_2 \ti Z_2$}

\subsection{ Differential Calculus on Discrete Groups $G$}
\hskip20pt Let $G$ be a discrete group of size $N_G$, its elements are
$\{e, g_1, g_2, \cdots , g_{N_G-1}\}$, and $\cal A$
the algebra of
the all complex valued functions on $G$ .The right and left multiplication
on $G$ induce natural automorphisms of $\cal A$, $R_g$ and $L_g$ respectively,
\be
(R_hf)(g)=f(g\cdot h),
(L_hf)(g)=f(h\cdot g) \ee
The basis $\partial_i, ~(i=1, \cdots , N_G-1) $
of the left invariant vector space $\cal F$ on $ \cal A$ are defined as
\be
\partial_if=f-R_if,~~\forall f\in \cal A, \ee
 $\partial_i$ also satisfies
\be
\partial_i \partial_j=\displaystyle\sum_k C_{ij}^k
\partial_k,~~~~ C_{ij}^k=\delta _i^k + \delta _j^k -
\delta _{i\cdot j }^k\ee
where  $i, j, \cdots, (i\cdot j)$ denote  $g_i, g_j, \cdots,
(g_i\cdot g_j)$ respectively.
The Haar integral, which remains invariant under group action, is introduced
as a complex valued linear functional on $\cal A$ as,
\be
\int_G f=\frac 1 {N_G}\sum_{g\in G}f(g).
\ee
Having chosen the basis of $\cal F$ we can introduce the dual basis of $\cal
F^*$ consisting of one forms, $\chi^i$, satisfy
\be
 \chi^i(\partial_j)=\delta^i_j. \ee
There exists exactly one linear operator $d, d:\O^n \rightarrow \O^{n+1}, $
which is nilpotent, $d^2=0$, satisfies the graded Leibniz rule and for every
$f\in \cal A $ and every vector field $v, df(v)= v(f)$, provided that $\chi^i$ 
satisfy the following two conditions\cite{Sit}
\be
\begin{array}{cl}
\chi^if=&(R_if)\chi^i,~~\forall f\in \cal A,\\[3mm]
d\chi^i=&-\sum_{j,k}C_{jk}^i\chi^j\otimes \chi^k.
\end{array}
\ee
To construct a physical model we must define the involution operator * on $A$ ,
which agrees with the complex conjugation on ${\cal A}$, takes the assumption
that $(\chi^g)^*=-\chi^{g^{-1}}$, and (graded) commutes with d, i.e. 
$d(\omega^*)=(-1)^{deg\omega}(d\omega)^*$.

\subsection{ Gauge Theory on Discrete Group $G$}
Using the differential calculus introduced in last section, we can
 construct the generalized gauge theory on finite groups. The gauge 
 transformations group is often taken to be the group of unitary 
 elements of zero-forms $\tilde{A}$,
\be {\cal H} ={ \cal U}(\tilde{A})= {a \in\tilde{A}:aa^*=a^*a=1}\ee
Like the usual gauge theory, the
 $d+\phi$ is gauge covariant which requires the transformation of
 gauge field one form $\phi$ as,
\be
\phi \rightarrow H\phi H^{-1}+HdH^{-1}. \ee
If we write $\phi =\displaystyle{\sum_g}\phi_g \chi^g$,
the coefficients $\phi_g$ transform as
\be
 \phi_g \rightarrow H\phi_g(R_gH^{-1})+H\partial_gH^{-1}.
\ee
It is convenient to introduce a new field $\Phi_g=1-\phi_g$, then
(2.10) is equivalent to
\be
\Phi_g \rightarrow H\Phi_g(R_gH^{-1}). \ee
The extended anti-hermitian condition $\phi^* = -\phi$ results in the
following relations on its coefficients $\phi_g$ as well as $\Phi_g$,
\be
\phi_g^{\dag} = R_g(\phi_{g^{-1}}), ~~~~\Phi_g^{\dag}=R_g(\Phi_{g^{-1}})
\ee
It can be easily shown that the curvature two form
{}~~$ F=d\phi+\phi \otimes \phi $~~
is gauge covariant and can be written in terms of its  coefficients
\be
 F=\displaystyle{\sum_{g,h}}F_{gh}\chi^g\otimes\chi^h \ee
\be
F_{gh}=\Phi_gR_g(\Phi_h)-\Phi_{h\cdot g}.\ee
In order to construct the Lagrangian of this gauge theory on
discrete groups
we need to introduce a metric on the forms. Let us first define the
metric $\eta$ as a bilinear form on the bimodule $\Omega^{1}$ valued in the
algebra ${\cal A}$,
\be
\eta : ~~\Omega^1 \otimes \Omega^1 \rightarrow \cal A \ee
\be
 <\chi^g, \chi^h>=\eta^{gh}.
 \ee
The gauge invariance requires that $\eta^{gh} \sim
\delta^{gh^{-1}}$\cite{Ding}. The metric on the two forms becomes,
\be
<\chi^g \otimes \chi^h, \chi^p \otimes \chi^q >=
\alpha \eta^{gh}\eta^{pq} + \beta \eta^{gq}\eta^{hp} +
\gamma \eta^{gp}\eta^{hq},\ee
where the term proportional to $\gamma$ is only appeared when $G$ is
commutative \cite{Sit}.In the next section ,We will get a constrain relation of 
$\a, \b $ and $\g $ if the spontaneous $CP$ violation is considered.  
Then the most general Yang- Mills action is given,
\be
\ba{cl}
 {\cal L}&=\displaystyle-\int_G <F,\overline F> \\[3mm]
&=\displaystyle-\int_G \sum_{g,h,p,q} Tr(F_{gh}F^{\dag}_{pq})
<\chi^g \otimes \chi^h, \chi^{g^{-1}}\otimes
\chi^{p^{-1}}>\\[3mm]
&=\displaystyle-\int_G \sum_{g,h,p,q} Tr(F_{gh}F^{\dag}_{pq})
(\alpha \eta^{gh}\eta^{q^{-1}p^{-1}}
+ \beta \eta^{gp^{-1}}\eta^{hq^{-1}} +
\gamma \eta^{gq^{-1}}\eta^{hp^{-1}}),
\ea \ee
where we have used the involution relations
\be
{\overline F}=(\chi^q)^* \otimes(\chi^p)^* F^{\dag}_{pq},~~~~~
(\chi^g)^*=-\chi^{g^{-1}}.\ee

\subsection{ Gauge Field Theory on $M^{4}\ti G$}

To deal with the model building in particle physics, we must include fermions
and their Yukawa couplings to Higgs. A detailed construction of the gauge 
theory on discrete groups coupled to the fermions has given in \cite{Ding} 
\cite{Li}. Here we
only site some conclusion which is useful to the model building in the next
section.  $M^{4}$ is the four-dimension space-time. We defined 
the exterior derivative operators $d_{M}$ and $d_{G}$ on $M^{4}$ 
and $G$ respectively,i.e $ d_M f= \pa_{\mu}f dx^{\mu}, ~~d_G f=\pa_{g}f 
\chi^{g}$
, then $d_{M^{4}\ti G}= d_M + d_G $ can be extended to the covariant derivative 
\be
\ba{ll}
D_{M^{4}\ti G}&=d_{M^{4}\ti G}+A\\
d_{M^{4}\ti G} f&=\pa_{\mu}f dx^{\mu}+ \pa_{r}f x^{r}\\
A &=igA_{\mu}dx^{\mu} + \frac{1}{\mu}\phi_{g} \chi x^{g}
\ea\ee
The generalized curative two form:
\be
\ba{ll}
F&=dA + A \otimes A\\
 &= \frac{ig}{2} F_{\mu\nu} dx^{\mu} \w dx^{\nu}+ \frac{1}{\mu} F_{g\mu} 
dx^{\mu} \otimes  \chi^{g}+ \frac{1}{\mu^2} F_{gh} \chi^{g} \otimes \chi^{h}
\ea\ee
where\\
\be
\ba{ll} F_{\mu\nu}&=\pa_{\nu} A_{\mu} -\pa_{\nu} A_{\mu} +ig[A_{\mu},A_{\nu}]\\
[2mm]
F_{g\mu}&=\pa_{\mu} \phi_{g}+igA_{\mu}\phi_{g} -ig\phi_{g}R_{g} A_{\mu}\\[2mm]
F_{gh}&=\Phi_gR_g(\Phi_h)-\Phi_{h\cdot g}. \ea\ee
In next section we will find that $ F_{\mu\nu}$ is the ordinary gauge field
intense, $ F_{g\mu}$ will lead to the kinetic term of Higgs, and $F_{gh}$ 
lead to the Higgs potential.

The "local gauge invariant" fermion field Lagrangian on $ M_4 \ti G$ is written 
as,\cite{Li}
\be
{\cal L}=\overline\psi(x,g)[i\g^{\mu}( \pa_{\mu}+
 A_{\mu}(x,g))+ \mu \displaystyle{\sum_{h}}(\pa_{h}+ \frac{1}{\mu}\phi_{h}
 (x,g)R_{h})] \psi(x,g) \ee

\section{The Georgi-Glashow-Lee Model on the Discrete Group $Z_2 \ti Z_2$}
This model is a generalization of the Georgi-Glashow model given by T.D. Lee.
\cite{Lee}
The basic gauge group of the weak and electromagnetic interaction is $SO_3$.
There is a triplet of spin 1 gauge field $W_{\mu}$ and at least eight quark-like
hadron fields grouped into two triplets $ \psi, \psi^{'}$, as well as two 
singlets $\chi, \chi^{'}$. 

These fields can be expressed by the physical states $ p^{+}, n^{0}, 
\lambda^{0}, q^{-}$
and $p^{+'}, n^{0'}, \lambda^{0'}, q^{-'}$ but here we focused on two $SO_3$ 
triplets
of spin 0 Hermitian field ${\vec{\Phi}}_R, {\vec{\Phi}}_I $, which play 
important roles in $CP$ spontaneous 
violation of this model, Under the time reversal transformation $T$,
\be T {\vec{\Phi}}_RT^{-1} = +{\vec{\Phi}}_R,~~~but~~~ T{\vec{\Phi}}_IT^{-1}= - 
{\vec{\Phi}}_I
\ee
The total Lagrangian density can be written as 
\be {\cal L} = 
{\cal L} (W, \psi) +{\cal L }({\vec{\Phi}}_R, \psi)
+{\cal L}({\vec{\Phi}}_I, \psi)+{\cal L} (W,\vec{\Phi}) \ee
in which ${\cal L} (W, \psi)$ contain massive free fermions terms and their 
coupling with gauge fields $W$, and 
\be {\cal L }({\vec{\Phi}}_R, \psi)= -( g_R \bar{\psi}\vec{I}\psi+ 
{g_R}^{'} \bar{\psi^{'}}\vec{I}\psi^{'})\cdot {\vec{\Phi}}_R \ee
 
\be {\cal L }({\vec{\Phi}}_I, \psi)= -i( g_I \bar{\psi}\vec{I}\psi+ 
{g_I}^{'} \bar{\psi^{'}}\vec{I}\psi^{'})\cdot {\vec{\Phi}}_I \ee
 where $g_R$ and $g_I$ are real by hermiticity, and $\vec{I}$ is the $3\ti3$ 
matrix representation of $SO_3$ generators. For simplicity, we omitted all 
other $CP$ invariant and $SO_3$ invariant couplings between $ \Phi_R $ and
the hadron fields. We also assume  $ \Phi_R $ not directly interacting with
the singlet fields $\chi$ and $\chi^{'}$.
\be
{\cal L} (W,\vec{\Phi})= -\frac{1}{4} {\vec{W}}_{\mu\nu}^2 - 
\frac{1}{2}{\pa_{\mu}{\vec{\Phi}}_R}^2- 
\frac{1}{2}{\pa_{\mu}{\vec{\Phi}}_I}^2- V(\Phi)\ee
where 
\be 
{\vec{W}}_{\mu\nu}=\pa_{\mu} {\vec{W}}_{\nu}- \pa_{\nu} {\vec{W}}_{\mu}
+e ( {\vec{W}}_{\mu}\ti {\vec{W}}_{\nu})         \ee
\be   
\pa_{\mu}\vec{\Phi}_i=(\pa_{\mu}+e{\vec{W}}_{\mu}\ti) \vec{\Phi}_i~~~~~~~(i=R,I 
) \ee
\be 
V(\Phi)= -\frac{1}{2}\l_R{\vec{\Phi}}_R^2
                   - \frac{1}{2}\l_I{\vec{\Phi}}_I^2 
                   +\frac{1}{4}A_R({\vec{\Phi}}_R^2)^2
                               +\frac{1}{4}A_I({\vec{\Phi}}_I^2)^2
                   +\frac{1}{4}B({\vec{\Phi}}_I\cdot {\vec{\Phi}}_R)^2 
                    -\frac{1}{4}C {\vec{\Phi}}_R^2{\vec{\Phi}}_I^2 
\ee
in which the constants $\l_R, \l_I, ... ,C $ are real by Hermiticity.
As in \cite{Lee}, to insure that  $ V(\Phi )$ has a minimum, we assume 
\be A_R>0, ~~~A_I>0~~~and ~~~ A_I A_R > \frac{1}{4} (B-C)^2 \ee
In order that the minimum of $ V( \Phi )$ occurs at the values of
$\Phi_R $ and $\Phi_I $ given by
\be <{\vec{\Phi}}_R>={\vec{\rho}}_R \ne 0, ~~~~   
~     <{\vec{\Phi}}_I>={\vec{\rho}}_I \ne 0 \ee
we assume 
\be A_I \l_R > \frac{1}{2} (B-C) \l_I, ~~~~~~~A_R \l_I > \frac{1}{2} (B-C) \l_R
\ee
In addition we require $C>0$ so that ${\vec{\rho}}_R \parallel {\vec{\rho}}_I$.
\vskip 2mm
Now we apply the gauge theory on discrete group discussed in last section
to the case of $G=Z_2 \ti Z_2$. $G$ is isomorphism to Klein group $K_4$ and
has four elements, denoted by 
$ \{e, a,b, ab\} $ or the tensor product forms of $ Z_2=\{e,r\}, r^2=e$,
\be\ba{lll} 
e&=e_1 \otimes e_2,\hspace{5ex}
&a=e_1 \otimes r_2,\\[3mm]
b&=r_1 \otimes e_2,
&ab=r_1 \otimes r_2. \ea\ee
The multiplication$~~ * ~~$is defined as ,
\be g*h = g_1 h_1 \otimes g_2 h_2, ~~~ g=g_1 \otimes g_2,~~ h=h_1 \otimes h_2
\ee
There are three steps in our approach to reconstruct the physical model. First
we give a classification for all the fields on  $ M_4 \ti G$ according to their 
transformation characteristics with respect to the discrete symmetry, more 
precisely, we can define the right displacement $R_g$ as follows,
\be R_g f= g_1 g_2 f g_2^{-1} g_1^{-1}, ~~~~g= g_1 \otimes g_2 \ee
\vskip 2mm
If we take the $g_1= (CPT)^2, g_2= CPCP=PCPC $, where $C,P,T $ denote electrical
charge conjugation, parity, and time reversal transformations respectively,
we get a classification of the fields on  $ M_4 \ti G$. Then we apply the 
differential calculus and gauge theory on  $ M_4 \ti G$ to this case, and 
choose a consistent arrangement for those fields. Finally we do calculation 
straightforward and compare our results with the physical model we discussed.

Following this way, we arrange the fermion fields as,
\be\ba{ll}
\psi(x,e)=\left(\ba{c}\psi_L\\ \psi_R\ea\right)&
\psi(x,a)=-\left(\ba{c}\psi_L\\ \psi_R\ea\right) \\[5mm]
\psi(x,b)=-\left(\ba{c}\psi_L\\ \psi_R\ea\right)&
\psi(x,ab)=\left(\ba{c}\psi_L\\ \psi_R\ea\right)\ea\ee
and the gauge fields as,
\be W_{\mu} (x,h )= W_{\mu} (x) \ee
We also take the following arrangement of Higgs fields,
\be\ba{ll}
\Phi_{a}(x,h)=\left(
\ba{cc}
0&g_{R} \vec{I}\cdot {\vec{\Phi}}_{R}\\
 g_{R} \vec{I}\cdot {\vec{\Phi}}_R&0\ea\right),&
\Phi_{b}(x,h)=\left(
\ba{cc}
0&{-ig_{I} \vec{I}\cdot {\vec{\Phi}}_I}\\
 ig_{I} \vec{I}\cdot {\vec{\Phi}}_I&0\ea\right),\\[2mm]
 \Phi_{e}(x,h)= 0, &\Phi_{ab}(x,h)=0
 \ea\ee
where $h$ is any element of $G$. It can be checked that these arrangements are
consistent and satisfy the extended anti-Hermitian condition (2.11).  According 
to the equations (2.22), the Yukawa coupling terms are,
\be \ba{ll}
{\cal L }_Yukawa &\propto \overline\psi(x,g) \phi_{h}(x,g)R_{h} \psi(x,g) 
\\[2mm]
 &\propto \overline\psi(x,g) \phi_{a}(x,g)R_{a} \psi(x,g)+ \overline\psi(x,g) 
\phi_{b}(x,g)R_{b} \psi(x,g) \\[2mm]
 &\propto -g_R \overline\psi_L \vec{I}\cdot {\vec{\Phi}}_R  \psi_R 
          -g_R \overline\psi_R \vec{I}\cdot {\vec{\Phi}}_R  \psi_L 
          -ig_I \overline\psi_L \vec{I}\cdot {\vec{\Phi}}_I  \psi_R 
          +ig_I \overline\psi_R \vec{I}\cdot {\vec{\Phi}}_I  \psi_L \\[2mm]
&\propto -g_R \overline\psi \vec{I}\cdot {\vec{\Phi}}_R  \psi 
          -ig_I \overline\psi \vec{I}\cdot {\vec{\Phi}}_I \g_5  \psi 
\ea\ee
where $  \g_5  \psi_R = + \psi_R,~~  \g_5  \psi_L = - \psi_L$ have been used.
For anther triplet $\psi^{'}, $ the similar procedure can be done.As mentioned 
in previous section, $F_{g\mu}$ lead to the Higgs kinetic terms
which is exactly the equation (3.29). It can be easily calculated because on 
$M_4 \ti G $  we have required the ordinary gauge fields keep the same. 
\vskip 2mm
Now we calculate the Higgs potential which is more complicated.  The only 
nontrivial curvatures are,
\be\ba{lll}
F_{aa}&=\Phi_{a}\cdot R_{a}\Phi_{a}-1,\hspace{5ex}
&F_{bb}=\Phi_{b}\cdot R_{b}\Phi_{b}-1,\\[3mm]
F_{ab}&=\Phi_{a}\cdot R_{a}\Phi_{b},
&F_{ba}=\Phi_{b}\cdot R_{b}\Phi_{a}. \ea\ee
\vskip 2mm
If we take the matrix representation of the $SO_3$ generators $\vec{I}$ 
and rewrite $\vec{I}\cdot {\vec{\Phi}}_R$ and $ \vec{I}\cdot {\vec{\Phi}}_I$
as follows,
\be\ba{ll}
\vec{I}\cdot {\vec{\Phi}}_R=\left(
\ba{ccc}
0&-i\Phi_{R3}&i\Phi_{R2}\\i\Phi_{R3}&0&-i\Phi_{R1}\\-i\Phi_{R2}
&i\Phi_{R1}&0\ea\right), &
\vec{I}\cdot {\vec{\Phi}}_I=\left(
\ba{ccc}
0&-i\Phi_{I3}&i\Phi_{I2}\\i\Phi_{I3}&0&-i\Phi_{I1}\\-i\Phi_{I2}
&i\Phi_{I1}&0\ea\right),\ea \ee
we have
\be\ba{ll}
Tr [(\vec{I}\cdot {\vec{\Phi}}_i)^2]=2{\vec{\Phi}}_i^2 &
Tr [(\vec{I}\cdot {\vec{\Phi}}_i)^4]=2({\vec{\Phi}}_i^2)^2~~~~~~~~~i=R, I 
\\[2mm]
Tr [((\vec{I}\cdot {\vec{\Phi}}_R)(\vec{I}\cdot {\vec{\Phi}}_I))^2]
=2({\vec{\Phi}}_R\cdot\vec\Phi_I)^2 &
 Tr [(\vec{I}\cdot {\vec{\Phi}}_R)^2(\vec{I}\cdot {\vec{\Phi}}_I)^2]
=({\vec{\Phi}}_R\cdot\vec\Phi_I)^2 +{\vec{\Phi}}_R^2{\vec{\Phi}}_I^2
 \ea \ee
With the help of these identities and the equation ( 2.17 ), the "Yang-Mills
term" can be calculated straightforward,
\be\ba{lll}
<F, \overline F>& = Tr[&\a(E_a^2 F_{aa}+2E_a E_b F_{aa} F_{bb}+E_b^2 F_{bb})\\
 [2mm]               &      &+\b(E_a^2 F_{aa}+2E_a E_b F_{ab} F_{ba}+E_b^2 
F_{bb})+ 
 \g(E_a^2 F_{aa}+2E_a E_b F_{ab} F_{ba}+E_b^2 F_{bb}\\[2mm]
                & = Tr[&(\a+\b+\g)(E_a^2 F_{aa}+E_b^2 F_{bb})+2(\g-\b)E_a E_b
F_{ab}^2+2\a E_a E_b F_{aa} F_{bb}]  \ea\ee
Then we have the following expression of Higgs potential similar with the one
in \cite {Lee}, i.e (3.30),
\be\ba{ll}
V({\vec{\Phi}}_R,{\vec{\Phi}}_I)= &2(\a+\b+\g)E_a^2 g_R^4 ({\vec{\Phi}}_R^2)^2
                               +2(\a+\b+\g)E_b^2 g_I^4 
({\vec{\Phi}}_I^2)^2\\[2mm]
                 & -4g_R^2[(\a+\b+\g)E_a^2+\a E_a E_b] {\vec{\Phi}}_R^2
                   -4g_I^2[(\a+\b+\g)E_b^2+\a E_a E_b] {\vec{\Phi}}_I^2 \\[2mm]
  &+[4(\g-\b)+2\a]E_a E_b g_R^2 g_I^2({\vec{\Phi}}_I\cdot {\vec{\Phi}}_R)^2+ 
  2\a E_a E_b g_R^2 g_I^2 {\vec{\Phi}}_R^2{\vec{\Phi}}_I^2 
 \ea\ee
If we absorbed the Yukawa coupling constants $ g_R , g_I $ into the Higgs fields 
, and require that,
\be
E_a > 0, ~~~~~ E_b > 0 \ee
\be
\a > 0, ~~~~\a+\b+\g > 0, ~~~~~\b-\g >  \frac{\a}{2}\ee
The condition (3.31)-(3.33) are satisfied automatically.

\section{Weinberg-Branco Model on the Discrete Group $Z_2 \ti Z_2 \ti Z_2$ }
It was shown that if one imposed NFC in the context of the standard $SU_2 \ti
U_1 $ gauge theory, then at least three Higgs doublets are needed in order to
violate $CP$. In \cite{Chen} a three-Higgs toy model of SCPV was constructed
via the gauge theory on  $Z_2 \ti Z_2 \ti Z_2$, which allow each Higgs couple
to  all the quarks but ensure the NFC by imposing some constrains on the
Yukawa coupling constant matrices. Here we choose another approach to ensure
the NFC, i.e 
preventing the third Higgs doublet $\Phi_3$ from coupling to quarks. As in 
\cite{Wein} 
, we take the element of third $Z_2$ group to be a reflection symmetry$R^{'}:
 \Phi_3 \rightarrow -\Phi_3$ while all other fields remain unchanged. The first
  $Z_2$ group is still generated by $ \theta= (CPT)^2$, the second $Z_2$ group 
is 
  generated by $ R: \Phi_2 \rightarrow -\Phi_2,~~D_R^i \rightarrow -D_R^i$ with
  all the other fields unchanged.
  
  Following the method in last section, we construct the Weinberg-Salam-Branco
  model on the discrete group $\theta \ti R \ti R^{'}={g_0,g_1,g_2,g_3,g_4,g_5,
  g_6,g_7}$
\be\ba{lll} 
g_0&=e_1 \otimes e_2 \otimes e_3,\hspace{5ex}
&g_1=r_1 \otimes e_2 \otimes e_3,\\[2mm]
g_2&=e_1 \otimes r_2 \otimes e_3,
&g_3=e_1 \otimes e_2 \otimes r_3,\\[2mm]
g_4&=r_1 \otimes r_2 \otimes e_3,
&g_5=r_1 \otimes e_2 \otimes r_3,\\[2mm]
g_6&=e_1 \otimes r_2 \otimes r_3,
&g_7=r_1 \otimes r_2 \otimes r_3.
\ea\ee
where $e_i$ denotes identity, $r_i$ denotes the transformation of each discrete 
group respectively. The multiplication$~~ * ~~$is defined as ,
\be g*h = g_1 h_1 \otimes g_2 h_2 \otimes g_3 h_3, 
~~~ g=g_1 \otimes g_2 \otimes g_3,~~ h=h_1 \otimes h_2 \otimes h_3 \ee
According to their transformation properties of the symmetry$ \theta, R $ and $ 
R^{'}$,
i.e.$R_g f= g_1 g_2 g_3f g_3^{-1} g_2^{-1} g_1^{-1}, g= g_1 \otimes g_2 \otimes 
g_3 $
 we take the following arrangement of the fermion fields , gauge fields and 
Higgs fields, 
( To compare with the model in \cite{Wein}, we only write down the quark fields.
the inclusion of lepton fields is straightforward and given in \cite{Li}  
\cite{Chen}.)
\be \psi(x,g)=\left(\ba{c}L(x,g)\\R(x,g)\ea\right),
~~L(x,g)=\left(\ba{c}U(x,g)\\D(x,g)\ea\right)_L,
~~R(x,g)=\left(\ba{c}U(x,g)\\D(x,g)\ea\right)_R, \ee
where $U=(u,c,t),~~D=(d,s,b)$, with the left-handed components $ U_L^i, D_L^i$ 
forming $SU_2$ doublets 
$ (U_L^i, D_L^i)$, while the right-handed components are singlets. 

\be\ba{llll}
U^i(x,g)= &+ U^i (x), ~if ~~g=g_0,g_2,g_3,g_6.~~&D_R^i(x,g)=&+D_R^i(x),~if 
~~g=g_0,g_3,g_4,g_7\\[2mm]
          &- U^i (x), ~if ~~g=g_1,g_4,g_5,g_7.~~&           &-D_R^i(x),~if 
~~g=g_1,g_2,g_5,g_8

\ea\ee  
and $ D_L^i(x,g)$ is arranged as same as $ U^i(x,g)$.  
and the gauge fields as,
\be
A_{\mu}(x,g)=A_{\mu}(x), ~~ for~~~ g \in G\ee
The Higgs fields,
\be\ba{ll}
\Phi_{g_1}(x,h)=&\left(
\ba{cc}
0& \Phi_1(x)\\
\Phi_1^{\dag}(x)&0\ea\right),~for~~~ g\in G\ea\ee
\be\ba{ll}\Phi_{g_3}(x,g)=&\left(
\ba{cc}
0&\Phi_2(x)\\ 
\Phi_2^{\dag}(x)&0\ea\right),~if~g=g_0,g_1,g_3,g_5 \\[3mm]
~~~~~~~~~~~or ~~-&\left(\ba{cc} 0& \Phi_2(x)\\
\Phi_2^{\dag}(x)&0\ea\right),~if~g=g_2,g_4,g_6,g_7  
\ea\ee
\be\ba{ll}\Phi_{g_4}(x,g)=&\left(
\ba{cc}
0&\Phi_3(x)\\ 
\Phi_3^{\dag}(x)&0\ea\right),~if~g=g_0,g_1,g_2,g_4 \\[3mm]
~~~~~~~~~~~or ~~-&\left(\ba{cc} 0& \Phi_3(x)\\
\Phi_3^{\dag}(x)&0\ea\right),~if~g=g_3,g_5,g_6,g_7  
\ea\ee
\vskip 2mm
\be 
\Phi_{g_0}(x,g)=\Phi_{g_2}(x,g)=  
\Phi_{g_5}(x,g)=\Phi_{g_6}(x,g)=\Phi_{g_7}(x,g)=0 \ee
where 
\be \Phi_n(x)=\left(\ba{ccc}
{\phi}_n^{0*}&{\phi}_n^{+}\\[1mm]
-{\phi}_n^{+*}&{\phi}_n^{0}\ea \right) \otimes I_i~.~\left(\ba{ccc}
{\G}_n^{U} & \\[1mm]
 &{\G}_n^{D}\ea \right) \ee
${\G}_n^{U},{\G}_n^{D} $ are the coefficients of Yukawa couplings.

The above arrangement of the quarks and Higgs fields differ from the arrangement 
in \cite{Chen}, because we took the second and the third $Z_2$ to be the $R$
and $R^{'}$ symmetry respectively. This modification
is important to prevent the third Higgs particle from coupling to the quarks, 
the Higgs $\Phi_1$ from coupling to $D_R^i$
as well as the Higgs $\Phi_2$ from coupling to $U_R^i$. We have
\be \ba{ll}
{\cal L }_{Yukawa} &= \displaystyle{\sum_{g}}\displaystyle{\sum_{h}}
\overline\psi(x,g) \phi_{h}(x,g)R_{h} \psi(x,g) 
\\[2mm]
&=-\displaystyle{\sum_{n=1,2}} 
\displaystyle{\G}^U_{2ij}({\bar{U}}^i_L{\p}^{0\ast}_2-
{\bar{D}}^i_L{\p}^{+\ast}_2)
U^j_R+{\G}^D_{1ij}({\bar{U}}^i_L{\p}^{+}_1+{\bar{D}}^i_L{\p}^0_1)D^j_R+ h.c.
\\[2mm]
&= {\G}^U_{2ij}{\bar L}^i {\pi}_2 D^j_R + 
{\G}^U_{1ij}{\bar L}^i \tilde{{\pi}_1} U^j_R + h.c.\ea\ee
where $
L^i= ({\bar{U}}^i_L,{\bar{D}}^i_L), \pi_n= {({\p}^{+}_n, {\p}^{0}_n)}^T$,and 
$\tilde{{\pi}_1}=i {\sigma}_2 {\pi}_1 $.

Thus we obtained the Yukawa interactions discussed in \cite{Wein}.
On the other hand, as we expected, this modification does not impact on the 
Higgs kinetic terms and the Higgs
potential, which still contain three Higgs as in \cite{Chen}
\be \ba{ll}
V(\Phi_1,\Phi_2,\Phi_3)=&\alpha Tr{[E_1(\Phi_1\Phi_1^{\dag}-1)+
E_2(\Phi_2\Phi_2^{\dag}-1)+E_3(\Phi_3\Phi_3^{\dag}-1)]}^2\\[2mm]
&+\beta
[E_1^2Tr{(\Phi_1\Phi_1^{\dag}-1)}^2+E_2^2Tr{(\Phi_2\Phi_2^{\dag}-1)}^2
+E_3^2Tr{(\Phi_3\Phi_3^{\dag}-1)}^2\\[2mm]
&+2E_1E_2Tr\Phi_1\Phi_2^{\dag}\Phi_2\Phi_1^{\dag}
+2E_1E_3Tr\Phi_1\Phi_3^{\dag}\Phi_3\Phi_1^{\dag}
+2E_3E_2Tr\Phi_3\Phi_2^{\dag}\Phi_2\Phi_3^{\dag}]\\[2mm]
&+\gamma [E_1^2Tr(\Phi_1\Phi_1^{\dag}-1)^2+E_2^2Tr(\Phi_2\Phi_2^{\dag}-1)^2+
E_3^2Tr{(\Phi_3\Phi_3^{\dag}-1)}^2 \\[2mm]
&+E_1E_2Tr(\Phi_1\Phi_2^{\dag}\Phi_1\Phi_2^{\dag}
+\Phi_2\Phi_1^{\dag}\Phi_2\Phi_1^{\dag})\\[2mm]
&+E_1E_3Tr(\Phi_1\Phi_3^{\dag}\Phi_1\Phi_3^{\dag}
+\Phi_3\Phi_1^{\dag}\Phi_3\Phi_1^{\dag})\\[2mm]
&+E_3E_2Tr(\Phi_3\Phi_2^{\dag}\Phi_3\Phi_2^{\dag}
+\Phi_2\Phi_3^{\dag}\Phi_2\Phi_3^{\dag})]\ea\ee
It is very important that the third Higgs isn't cancelled from the Higgs  
potential,
for at least three Higgs are needed to violate $CP$ in the Weinberg-Salam-Branco
model.
\be\ba{ll}
Tr(\Phi_m\Phi_m^{\dag})~~~~~~~=&
2Tr({\G}^U_m{\G}^{U\da}_m+{\G}^D_m{\G}^{D\da}_m)(\pi_m^{\dag}\pi_m)\\[2mm]
Tr(\Phi_m\Phi_m^{\dag}\Phi_m\Phi_m^{\dag})=&
2Tr({\G}^U_m{\G}^{U\da}_m{\G}^U_m{\G}^{U\da}_m+{\G}^D_m{\G}^{D\da}_m{\G}^D_m{\G}
^{D\da}_m)
(\pi_m^{\dag}\pi_m)(\pi_m^{\dag}\pi_m)\\[2mm]
Tr(\Phi_m\Phi_n^{\dag}\Phi_m\Phi_n^{\dag})=&
2Tr({\G}^U_m{\G}^{U\da}_n{\G}^U_m{\G}^{U\da}_n) 
(\pi_n^{\dag}\pi_m)(\pi_n^{\dag}\pi_m)+
2Tr({\G}^D_m{\G}^{D\da}_n{\G}^D_m{\G}^{D\da}_n) 
(\pi_m^{\dag}\pi_n)(\pi_m^{\dag}\pi_n)
\\[2mm]
&+4Tr({\G}^D_m{\G}^{D\da}_n{\G}^U_m{\G}^{U\da}_n)
[(\pi_m^{\dag}\pi_n)(\pi_n^{\dag}\pi_m)-(\pi_m^{\dag}\pi_m)(\pi_n^{\dag}\pi_n)]
\\[2mm]
Tr(\Phi_m\Phi_n^{\dag}\Phi_n\Phi_m^{\dag})=&
2Tr({\G}^U_m{\G}^{U\da}_n{\G}^U_n{\G}^{U\da}_m+{\G}^D_m{\G}^{D\da}_n{\G}^D_n{\G}
^{D\da}_m) 
(\pi_m^{\dag}\pi_m)(\pi_n^{\dag}\pi_n)\\[2mm]
Tr(\Phi_m\Phi_m^{\dag}\Phi_n\Phi_n^{\dag})=&
2Tr[({\G}^U_m{\G}^{U\da}_m+{\G}^D_m{\G}^{D\da}_m)]
({\G}^U_n{\G}^{U\da}_n+{\G}^D_n{\G}^{D\da}_n)(\pi_m^{\dag}\pi_n)(\pi_n^{\dag}\pi
_m)
\\[2mm]
&-2Tr({\G}^D_m{\G}^{D\da}_m{\G}^U_n{\G}^{U\da}_n+{\G}^U_m{\G}^{U\da}_m{\G}^D_n{\
G}^{D\da}_n)
(\pi_m^{\dag}\pi_m)(\pi_n^{\dag}\pi_n)\ea\ee
With the help of these identities, we obtained
\be \ba{cl}
V(\pi_1,\pi_2,\pi_3)&=\displaystyle\sum_{n=1}^3[a_{nn}(\pi_n^{\dag}\pi_n)^2\pi_n
+
m_n(\pi_n^{\dag}\pi_n)]
+\displaystyle\sum_{n<m}\{a_{nm}(\pi_n^{\dag}\pi_n)(\pi_m^{\dag}\pi_m)\\[3mm]
&+b_{nm}(\pi_n^{\dag}\pi_m)(\pi_m^{\dag}\pi_n)
+[c_{nm}(\pi_n^{\dag}\pi_m)(\pi_n^{\dag}\pi_m)+h.c.]\}\ea\ee
where \be c_{nm} = 2\g E_n E_m Tr({\G}^U_m{\G}^{U\da}_n{\G}^U_m{\G}^{U\da}_n+
{\G}^D_n{\G}^{D\da}_m{\G}^D_n{\G}^{D\da}_m)\ee
\vskip 2mm
 If we assume the minimum to be at
$ \pi_i=\f {1}{\sqrt{2}} ( 0, v_i e^{i{\theta}_i})^T $ and set $ \theta_2=0$,
we can classify the solutions of stationarity condition \cite{Des} as in  
\cite{Wein}

(1)$CP$- conserving solution:
\be \theta_1=\f {1}{2}n\pi,~~\theta_3=\f {1}{2}m\pi,~~m,n\in \cal{Z} \ee

(2)$CP$- violating solution:
\be cos2\theta_1 = \f {1}{2} \left( \f{d_{13}d_{23}}{d_{12}^2}- 
\f{d_{23}}{d_{13}}
-\f{d_{13}}{d_{23}}\right),~~~~~
cos2\theta_3=\f {1}{2} \left( \f{d_{13}d_{12}}{d_{23}^2}-\f{d_{12}}{d_{13}}-
\f{d_{13}}{d_{12}}\right) \ee
where $ d_{nm}=c_{nm} v_n^2 v_m^2 $, $c_{nm}$ are given by (4.62). 

\section{conclusion and remark}
 We have reconstructed Georgi-Glashow-Lee model and
 Weinberg-Salam-Branco model with SCPV, based on the differential 
calculus and gauge theory on discrete group. We showed that in these models, all
the Lagrangian terms containing Higgs particles are related as a whole by the 
gauge 
theory with respect to certain discrete
symmetry, which can determine the Yukawa couplings, the Higgs kinetic terms and
the Higgs potential with less ambiguity. In the first model to insure that 
$V(\Phi)$ has
a minimum and the minimum occurs at the vacuum expected value of $\Phi_R,\Phi_I$
, the metric parameters must satisfy some condition. In the second model, to 
insure
NFC we take the two $Z_2$ symmetries to be the $R$ and $R^{'}$, find this 
modification also lead to SCPV.  
Thus we established a 
correlation between the physics model and the nontrivial geometry structure 
behind it. 
The constraints of NCG on the physics models are to be investigated further.

\vskip 1.5cm
{\large\bf {Acknowledgments}}\\
\vskip 2mm
One of the authors (C.Xiong) would like to thank Doctor Bin Chen for his lecture
on NCG. The helpful discussions with Professors Chao-Hsi chang and  Yue-Liang Wu
, Doctors Jianming Li and Wei Zhang are truly appreciated.

\end{document}